\documentclass[preprint,12pt]{elsarticle}

\usepackage{epsfig}
\usepackage{graphicx}

\newtheorem{theorem}{Theorem}
\newtheorem{lemma}[theorem]{Lemma}
\newtheorem{corollary}[theorem]{Corollary}
\newtheorem{proposition}[theorem]{Proposition}

\usepackage{geometry} 
\usepackage{amsmath}
\usepackage{amsfonts}
\usepackage{graphicx}
\usepackage{epsfig}
\usepackage{latexsym}
\usepackage{epsfig}
\newcommand{\PP}{{\mathbb P}}
\newcommand{\EE}{{\mathbb E}}
\newcommand{\ZZ}{{\mathbb Z}}
\newcommand{\A}{{\mathcal A}}

\journal{}

\begin{document}
\begin{frontmatter}

\title{The `Butterfly effect' in  Cayley graphs,
and its relevance for evolutionary genomics}
\author{Vincent Moulton and Mike Steel}
\address{(VM) School of Computing Sciences,
University of East Anglia,
Norwich, NR4 7TJ, UK. 
(MS) Biomathematics Research Centre, University of Canterbury, Christchurch, New Zealand.\\
email: vincent.moulton@cmp.uea.ac.uk , mike.steel@canterbury.ac.nz}
\begin{keyword}
evolutionary distance, permutation, metric, group action, genome rearrangements
\end{keyword}

\begin{abstract}
Suppose a finite set $X$ is repeatedly transformed by a sequence  of permutations of a certain type acting on an initial element $x$ to produce a final state $y$.
 We investigate how `different' the resulting state $y'$ to $y$ can be  if a slight change is made to the sequence, either by deleting
one permutation, or replacing it with another.  Here the `difference' between $y$ and $y'$ might be measured by
the minimum number of permutations of the permitted type required to transform $y$ to $y'$, or by some other metric.   We discuss this first in the general setting of sensitivity to perturbation  of walks in  Cayley graphs of  groups with a specified set of generators.  We then investigate some permutation groups and generators arising in computational genomics, and the statistical implications of
the findings.
\end{abstract}
\end{frontmatter}


\section{Introduction}

In evolutionary genomics, 
two genomes\footnote{For the purposes of this paper a {\em genome} 
is simply  an ordered sequence of objects -- usually taken from
the DNA alphabet or a collection of genes --  which may occur with or 
without repetition, and with or without an orientation (+,-).} are frequently 
compared by the minimum number of
`rearrangements' (of various types) required to transform one genome
into another \cite{F09}. This minimum number is then used 
to estimate of the actual number
of events and thereby the `evolutionary distance' between the 
species involved.   
Since both the precise number and the actual rearrangement 
events that occurred in
the evolution of the two genomes from a common ancestor are
unknown, it is pertinent to have some idea of how sensitive 
this distance
estimate might be to the sequence of events (not just 
the number) that really took place \cite{SM08}.  

This question has important implications for the 
accurate inference of evolutionary relationships between 
species from their genomes, and we discuss some of these further 
in Section~\ref{stats}. However, we 
begin by framing the type of mathematical questions 
that we will be considering in a general algebraic context. 

Let $G$ be a finite group,  whose
identity element we write as $1_G$,  and let $S$ be a subset of
generators, that is {\em symmetric} (i.e. closed under 
inverses, so $x \in S \Rightarrow x^{-1} \in S$).
In addition, let $\Gamma = Cay(G,S)$ be the associated {\em Cayley graph},
with vertex set $G$ and an edge connecting $g$ and $g'$ 
if there exists $s \in S$ with
$g'= gs$ (unless otherwise stated, we use the convention 
of multiplying group elements from left to right).
For any two elements $g,g' \in G$, the distance $d_S(g,g')$ in 
$Cay(G,S)$  is the minimum value of
$k$ for which there exist elements $s_1, \ldots, s_k$ of $S$ 
so that $g' = g s_1 \cdots s_k$ (for $g=g'$, we set $d_S(g,g')=0$). 
Note that $d_S$ is a metric, in particular, $d_S(g,g') = d_S(g',g)$, since
$S$ is symmetric. 

In this paper, our focus is on the following two quantities:
$$
\lambda_1(G,S): =  
\mathop{\displaystyle \max }_{\scriptscriptstyle g \in G, s \in S}
\limits^{} \{d_S(sg, g)\},
$$
 \mbox{ and }
$$
\lambda_2(G,S): = 
\mathop{\displaystyle \max }_{\scriptscriptstyle g \in G, s,s' \in S}
\limits^{} \{d_S(sg, s'g)\}.
$$

One way to view these quantities is via the following 
result which is easily proved.

\begin{lemma}
\label{interpret}
\mbox{ }
Let $S$ be a symmetric set of generators for a finite group $G$. Then:
\begin{itemize}
\item
$\lambda_1(G,S)$ is the maximum value of $d_S(g,g')$ 
between any pair of elements $g$ and $g'$ of $G$ for which 
$g=s_1 s_2 \cdots s_k$, and $g'= s'_1 s'_2 \cdots s'_k$,  
where $s'_i = s_i \in S$ for all
but at most one value (say $j$) for $i$, and $s'_j = 1_G$.

\item
$\lambda_2(G,S)$ is the maximum value of $d_S(g,g')$ 
between any pair of elements $g$ and $g'$ of $G$ for which 
$g=s_1 s_2 \cdots s_k$ and $g'= s'_1 s'_2 \cdots s'_k$  
where $s'_i = s_i \in S$ for all
but at most one value (say $j$) for $i$, and $s'_j \in S, s'_j \neq s_j$.
\end{itemize}
\end{lemma}

Thus,  $\lambda_1(G,S)$ tells us how much (under $d_S$) a product
of generators can change if we drop one value of $s$, whilst 
$\lambda_2(G,S)$ tells us how much (again under $d_S$) a product
of generators in $S$ can change if we substitute one 
value of $s$ by another $s'$
(see Fig. 1 for an example where $\lambda_2(G,S)=6$). 

\begin{figure}[t]
\label{figure0}\begin{center}
 \resizebox{8cm}{!}{
 \includegraphics{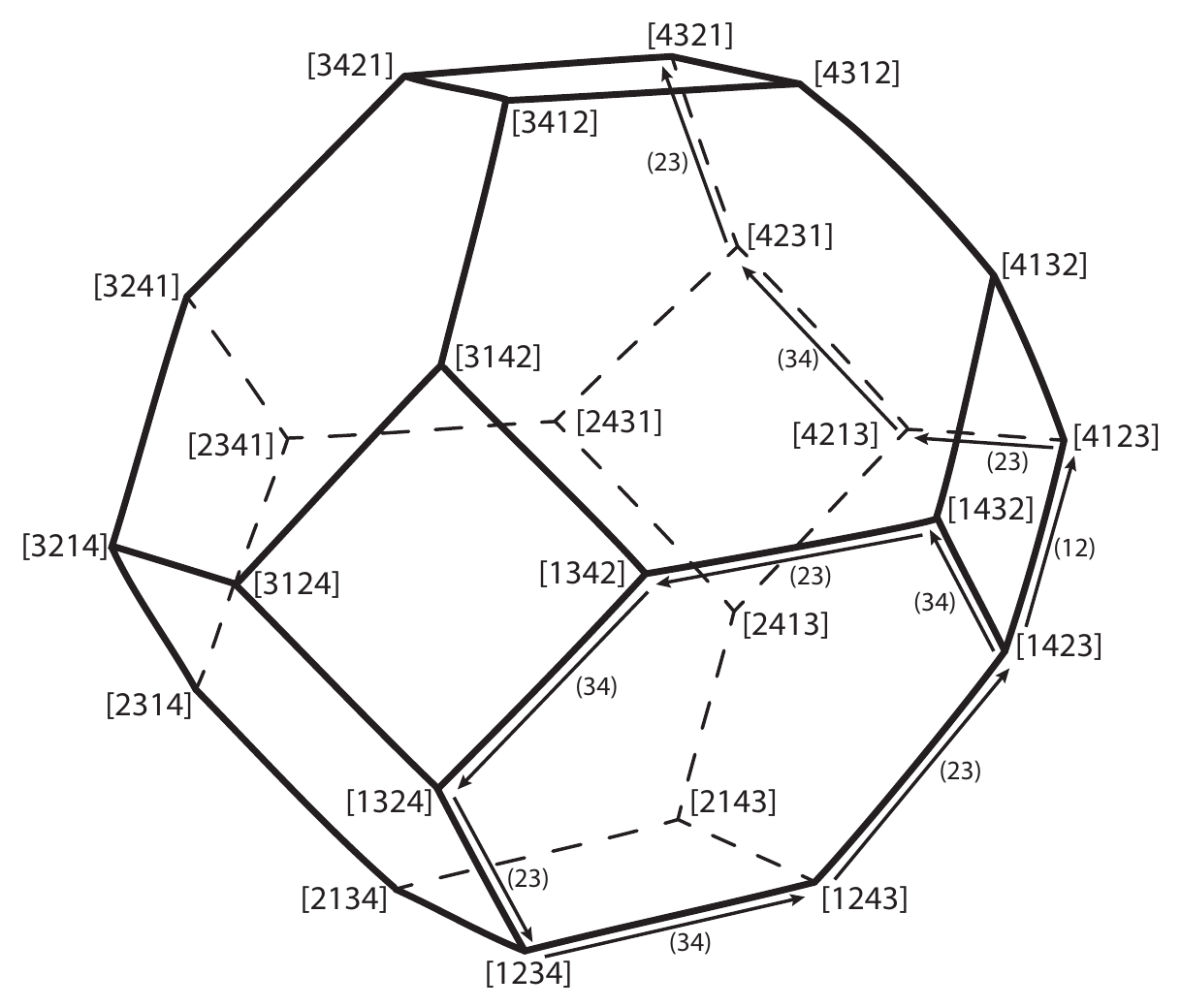}
 }
\end{center}
\caption{
The Cayley graph $Cay(G,S)$ for $G= \Sigma_4$ (the 
permutation group on $\{1,2,3,4\}$) 
and the set of transpositions $S=\{(12), (23), (34)\}$.  
Substituting just one element -- namely  (34) for (12) -- in 
the product corresponding to the walk in the lower front 
face (which starts and returns to the lower-most point
[1234]) results in a walk that ends at a point 
([4321], top) that is very distant (under $d_S$) from 
the end-point of the original walk. In fact, the two end-points are at 
maximal distance in this example.}
\end{figure}

As such, $\lambda_m$ is a measure of the `sensitivity' 
of walks in the Cayley graph
to a  switch in or deletion of a generator at some point.
Moreover, if $G$ acts transitively and freely\footnote{$G$ 
acts {\em transitively} on $X$ if for any pair $x,y \in X$ 
there exists $g \in G$ with 
$g \circ x = y$; the action is {\em free}  if 
$g \circ x = h \circ x \Rightarrow g=h$, for all $g,h \in G$ 
and $x \in X$, where  `$\,\circ\,$' denotes
the action of $G$ on $X$.}  on a set $X$ then $\lambda_m$ 
provides a corresponding 
measure of sensitivity of this action to a switch in 
or deletion of a generator 
(since a transitive, free action of $G$ on $X$ is isomorphic
to the action of $G$ on itself by right multiplication).  
Actions with 
large $\lambda_m$ values can thus be viewed as exhibiting a 
discrete, group-theoretic analogue of
the `butterfly effect' in non-linear 
dynamics (see e.g. \cite{Hilborn}).

In the genomics applications that we shall consider, 
elements of the group $G$ correspond to genomes, and $d_S$ to 
the evolutionary distance between them.
After presenting some general results concerning $\lambda_m$
in the next section, in Sections~\ref{genome} and \ref{break} 
we discuss some applications arising for 
various choices of $G$ and $S$. These include the
Klein four group, which arises in evolutionary models of 
DNA sequence evolution, and the permutatation
group, which typically appears when studying
rearrangement distances between genomes.  
We conclude in Section~\ref{stats} with 
some statistical implications of our results.

One can imagine many other settings besides genomics where similar
questions arise -- for example, in a sequence of moves that should
unscramble the Rubik's cube from a given position \cite{kunkle}, 
what will be the consequences (in terms of the
number of moves required) for completing the unscrambling if a mistake is
made at some point (or one move is forgotten)? 
In addition, related questions arise in the study of 
`automatic' groups, where the group under consideration
is typically infinite \cite{epp}.

\section{General inequalities}
\label{basec}

We first make some basic observations about Cayley graphs and the metric $d_S$ 
(further background on basic group theory, Cayley graphs, 
and group actions can be found in \cite{group}).
It is well known that $\Gamma$ is a connected
regular graph of degree equal to the cardinality of $S$ and that $\Gamma$
is also vertex-transitive (see, for example, \cite{cay}, Proposition 1). Consider the  function $l_S: G \rightarrow \{0, 1,2,3\ldots |G|\}$, where, $l_S(1_G) =0$ and, for each $g \in G-\{1_G\}$,
$l_S(g)$ is the smallest number $l$ of elements $s_1, \ldots, s_l$ from $S$ for which we can write
$g =s_1 \cdots s_l.$
The function $l_S$ clearly satisfies the subadditivity property that, for all $g,g' \in G$:
$$l_S(gg') \leq l_S(g)+ l_S(g').$$
In addition, $$l_S(g^{-1}) = l_S(g),$$
and
$$l_S(g)=1 \Leftrightarrow g \in S, l_S(g) = 0 \Leftrightarrow g = 1_G.$$
Note that $l_S(gg')$ is generally not equal to $l_S(g'g)$.
The metric $d_S$, described in the previous section, is related to $l_S$ as follows:
$$d_S(g,g') = l_S(g^{-1}g').$$
Consequently, by definition:

\begin{equation}
\label{l1eq}
\lambda_1(G,S) =  \mathop{\displaystyle \max }_{\scriptscriptstyle g \in G, s \in S}\limits^{} \{l_S(g^{-1}sg)\},
 \end{equation}
and
\begin{equation}
\label{l2eq}
 \lambda_2(G,S)= \mathop{\displaystyle \max }_{\scriptscriptstyle g \in G, s, s'  \in S}\limits^{} \{l_S(g^{-1}ss'g)\}.
 \end{equation}

 Let $l_S(G) = \max\{l_S(g): g \in G\}$, which is the diameter of $Cay(G,S)$,
that is, maximum length shortest path connecting any two elements of $G$.
 Clearly, $\lambda_1(G,S), \lambda_2(G,S)   \leq l_S(G)$.
Moreover:
\begin{equation}
\label{lemlam}
 \lambda_2(G,S)   \leq 2\cdot \lambda_1(G,S),
\end{equation}
since, for any $g \in G$ and $s,s' \in S$, we have:
$$d_S(sg, s'g) \leq d_S(sg, g) + d_S(g, s'g).$$
A partial converse to Inequality (\ref{lemlam}) is provided by the following:
\begin{equation}
\label{lemlam2}
\lambda_1(G,S) \leq \lambda_2(G,S) + \lambda'_1(G,S),
\end{equation}
where $\lambda'_1(G,S) =\max_{g \in G} \min_{s \in S} \{l_S(g^{-1}sg)\}.$
To verify (\ref{lemlam2}), select a pair $g\in G,s\in S$ so that $l_S(g^{-1}sg) =\lambda_1(G,S).$  Then:
$$\lambda_1(G,S)  = d_S(sg, g) \leq d_S(sg, s_1g) + d_S(s_1g, g),$$
where $s_1$ is an element $s'$ 
(possibly equal to $s$) in $S$ that minimizes $l_S(g^{-1}s'g)$.
Now, $d_S(sg, s_1g) \leq \lambda_2(G,S)$ (even if $s'=s$) and $d_S(s_1g, g) \leq \lambda'_1(G,S)$, and so we obtain
(\ref{lemlam2}).

Note also that if $G$ is Abelian, then $\lambda_1(G,S) =1$, and $\lambda_2(G,S) \leq 2$ for {\em any} symmetric set $S$ of generators.
Moreover, for the  Abelian 2-group $G = \ZZ_2^n$ and 
with  the symmetric set $S$ of
generators consisting of all $n$ elements with the identity at all but one position, we have
$l_S(G)=n$ and $\lambda_1(G,S)=1$. This shows that
the inequality $\lambda_1(G,S) \leq l_S(G)$  can be arbitrarily large.  Our next result generalizes this observation further.

\begin{lemma}
\label{baslem}
\mbox{  }
Let $G_1, G_2, \ldots, G_k$ be finite groups, and  let $S_i$ be a symmetric set of
generators of $G_i$ for $i=1,\ldots, n$. Consider the direct product 
$G = G_1 \times G_2 \times \cdots \times G_k$ along with the
symmetric set of generators $S$ of $G$ consisting of all possible $k$--tuples
which consist of the identity element of $G_i$ at all but one co-ordinate $i$,
where it takes some value in $S_i$. Then (i)
$\lambda_1(G, S) \leq   \max_{1 \leq i \leq k}\big\{l_{S_i}(G_i)\big\},$ and
(ii) $l_S(G) =  \sum_{i=1}^k l_{S_i}(G_i).$
\end{lemma}

{\em Proof:}
For Part (i), let $\lambda_1(G,S) = l_S(g^{-1}sg)$, where $s \in S$ is a non-identity element at some
co-ordinate $\nu$.  Notice that
$(g^{-1}sg)_j = 1_{G_j}$ for all $j \neq \nu$. Moreover,
$(g^{-1}sg)_\nu=   s_1\cdots s_l$ where $l \leq l_{S_\nu}(G_\nu)$.
Thus $l_S(g^{-1}sg) \leq  l_{S_\nu}(G_\nu)$, as claimed.

For Part (ii), the inequality  $l_S(G) \leq  \sum_{i=1}^k l_{S_i}(G_i)$ is clear; to establish the reverse inequality,
let $g_i$ be an element of $G_i$ with $l_{S_i}(g_i) = l_{S_i}(G_i)$, and $g = (g_1, \ldots, g_k) \in G$.  Then
$l_S(g)=   \sum_{i=1}^k l_{S_i}(G_i),$
and so $l_S(G) \geq  \sum_{i=1}^k l_{S_i}(G_i).$ 
\hfill$\Box$

We now consider how $\lambda_m$ behaves under group homomorphisms. 
Suppose $H$ is the homomorphic image of a group $G$ under a map $p$. Let $N = Ker(p)$ be
the kernel of $p$,  which is a normal subgroup of $G$, and with $H \cong  G/N$.  Thus we have a short exact sequence:

\begin{equation}
\label{exasec}
1 \rightarrow N \rightarrow G   \buildrel{p}\over{\rightarrow} H \rightarrow 1.
\end{equation}

Let $S$ be a symmetric set of generators of $G$. Then  $S_H= \{p(s): s \in S - N\}$ is  a symmetric set of generators of $H$.

\begin{lemma}
\label{lower}
For $m=1,2$, $\lambda_m(H, S_H) \leq \lambda_m(G, S).$
\end{lemma} 
{\em Proof:}
First suppose that $m=1$.
For $x \in S_H$ and $h \in H$, consider $h^{-1}xh$.  There exist elements $g \in G$ and $s \in S-N$ for which 
$f(g)=h$ and $f(s) = x$.  Now the element $g^{-1}sg \in G$ can be written as a product of at most $l = \lambda_1(G,S)$ elements of
$S$, that is $g^{-1}sg = s_1s_2 \cdots s_k$ for $k \leq l$.
Applying $p$ to both sides of this equation gives:
$h^{-1}xh = p(s_1)p(s_2) \cdots p(s_k)$.
Notice that some of the elements on right may equal the identity element of $H$ (since $p(s_i)=1_H \Leftrightarrow s_i \in N$), but they are elements of $S_H$ otherwise. Thus $l_{S_H}(h^{-1}xh) \leq l$.  Since this holds for all such elements $h,x$, Eqn. (\ref{l1eq}) shows that
$\lambda_1(H, S_H) \leq \lambda_1(G,S)$.  The corresponding result for $m=2$ follows by an analogous argument. 
\hfill$\Box$

To obtain a lower bound for $\lambda_m(G,S)$ suppose that the
short exact sequence (\ref{exasec}) is a {\em split extension}, i.e. there is a homomorphism $i: H \rightarrow G$ so that $p \circ i$ is the identity map on $H$, which (by the splitting lemma) is equivalent to the condition that $G$ is the semidirect product of $N$ with a subgroup $H'$ isomorphic to $H$ (i.e. $G = NH' =H'N, H'\cap N =\{1_G\}$). In this case we have the following bounds. 

\begin{proposition}
\label{semidirect}
Suppose a finite group $G$ is a semidirect product
of subgroups $N$ (normal) and $H$.
 Let $S_N, S_H$ be symmetric generator sets for $N$ and $H$ respectively, and let
$S= S_N \cup S_H$ which is a symmetric generator set for $G$.  Then:
$$
\lambda_1(H, S_H) \leq \lambda_1(G,S) \leq \lambda_1(H, S_H) + l_{S_N}(N).
$$
In particular,  by (\ref{lemlam}), $\lambda_2(G,S) \leq 2\lambda_1(H, S_H) + 2l_{S_N}(N).$
\end{proposition}

{\em Proof:}
The lower bound on $\lambda_1(G,S)$ follows from Lemma \ref{lower}.  For the upper bound we  must show that for all $s \in S$ and $g \in G$, $d_S(sg, g) \leq \lambda_1(H, S_H) + l_{S_N}(N)$ holds.
We consider two cases: (i) $s \in N$, and (ii) $s \in H$.
In Case (i), note that the conjugate element $g^{-1}sg$ is also an
element of $N$; in this case we have the tighter
bound $d_S(sg, g) \leq l_{S_N}(N)$.
In Case (ii), write $g=hn$ where $n \in N$ and $h \in H$. Consider the
word $$w = g^{-1}sg = n^{-1}h^{-1}shn.$$ Since  $N$ is normal
we have $n^{-1}(h^{-1}sh) = (h^{-1}sh)n'$ for some element $n' \in N$.
Thus $w= h^{-1}sh n'n.$ Write $w=w_1w_2$ where $w_1 = h^{-1}sh \in H$ and
$w_2 = n'n \in N$.  We can select $w_2$ to be a product of terms of $S_N$
of length at most $l_{S_N}(N)$ and, by Inequality (\ref{lemlam}),  we can select $w_1$ to
be a product of terms of $S_H$ of length at most $\lambda_1(H, S_H)$.
Thus $w$ can be written as a product of, at most,  $\lambda_1(H, S_H) + l_{S_N}(N)$
elements of $S$.
\hfill$\Box$

\section{Permutation groups and genomic applications}
\label{genome}

We first describe a direct application that is relevant to the evolution of a DNA sequence under a simple model of site substitution (Kimura's 3ST model) \cite{kim}.  Consider the four-letter DNA alphabet
$\A = \{A, C,G,T\}$ and the Klein four-group $K= \ZZ_2 \times \ZZ_2$ 
with an action on $\A$ in which 
the three non-zero elements of $K$ correspond to `transitions' (A $\leftrightarrow$ G, C $\leftrightarrow$ T)  and the two types of `transversions' (A$\leftrightarrow$C, G$\leftrightarrow$T; and A$\leftrightarrow$T, G$\leftrightarrow$C). This representation of the Kimura 3ST model  was first described and exploited by \cite{eva}. 

For $g \in K$ and $x \in \A$, let $g\circ x$ denote the element of $\A$ obtained by the action of $g$ on $x$ (the identity element fixes each element of $\A$). 
The resulting component-wise action of $K^n$ on $\A^n$, defined by:
$(g_1, \ldots g_n) \circ (x_1, \ldots x_n) = (g_1\circ x_1, \ldots, g_n \circ x_n)$,
can be regarded as the set of all changes that can occur to a  DNA sequence over a period of time under site substitutions.

Now, under any continuous-time Markovian process these change events (`site substitutions') occur just one at a time and so a natural generating set of $K^n$ is the set $S_n$ of all elements of $K^n$ that consist of $1_K$ at all but one co-ordinate.   Moreover, since the action of $K^n$ on $\A^n$ is  transitive and free (and so is isomorphic to the action of $K^n$ on itself by right multiplication),  $\lambda_m(K^n, S_n)$ measures the impact of  ignoring (for $m=1$) or replacing (for $m=2$) one substitution in a chain of such events over time. As $K^n$ is Abelian,  one has $\lambda_1(K^n, S_n)=1$ and $\lambda_2(K^n, S_n)=2$, which implies that this  impact is 
minor, and, more significantly, is independent of $n$; this has important statistical implications which we will describe further in Section \ref{stats}.

For a related example, consider the ordered sequence of distinct genes $(g_1, g_2,  \ldots, g_n)$ partitioned into regions $R_1, R_2, \ldots R_k$ so that genomic rearrangements occur within each region, but not between regions (e.g. 
$R_i$ might refer to different chromosomes).  This situation can be modelled by the setting of Lemma~\ref{baslem} in which
$G_i$ is a permutation group on the genes within $R_i$, and $S_i$ is set of elementary gene order rearrangement events that generates $G_i$ (we discuss some examples below). In this case, Lemma~\ref{baslem} provides a bound on $\lambda_1$ and
$\lambda_2$ that is independent of the number of regions $k$.

We turn now to the calculation of $\lambda_m(\Sigma_n,S)$ for the
permutation group $\Sigma_n$ on $n!$ elements and 
various sets $S$ of generators. This group 
commonly arises when studying genome rearrangements \cite{cay}.
Our main interest is to determine, for each instance of $S$,
whether there is a constant $C$ (independent of $n$) for which
$\lambda_m(\Sigma_n,S) \leq C$, for $m=1,2$.

A {\em permutation}  $g$ on the set $[n] := \{1,2,\dots,n\}$ is a
bijective mapping from $[n]$ to itself. We will also write $g$
as $g =[g_1,g_2,\dots,g_n]$ where $g_i = g(i)$ is the
image of the map $g$ for $i \in [n]$. 
Note that, following the usual convention,
the product $g g'$ of two permutations $g, g' \in \Sigma_n$ will
be considered as the composition of the functions $g$ and $g'$. In
particular, $gg'(i) = g(g'(i))$ for all $i \in [n]$.  

When studying genomes, each entry $g_i$ of a permutation $g$ 
corresponds to a gene and the full list
$[g_1,g_2,\dots,g_n]$ to a genome.
Multiplying $g$ by a permutation leads to a rearrangement of the 
genome. For example, multiplying by a 
{\em transposition}  $t_{i,j}$ interchanges the values at positions
$i$ and $j$ of $g$, 
i.e. $[\dots,g_i,\dots,g_j,\dots] t_{i,j} 
=[\dots,g_j,\dots,g_i,\dots]$, and multiplying by a {\em reversal} $r_{i,j}$
reverses the segment $[g_i,g_j]$, $1 \le i < j \le n$, of $g$, i.e.
$$
[\dots,g_i,g_{i+1}, \dots,g_{j-1}, g_j,\dots] r_{i,j} =
[\dots,g_j,g_{j-1}, \dots,g_{i+1}, g_i,\dots].
$$ 
Such rearrangements are widely observed and studied in
molecular biology \cite{F09}. 

In genomics applications, we are often interested
in defining some distance between genomes.  
One distance that is commonly used in 
the context of permutations is the {\em breakpoint}
distance \cite[7.3]{SM97}. For $g,g' \in \Sigma_n$,  
$d_{BP}(g,g')$ is defined as the number of pairs 
of elements that are adjacent in the list $[0,g_1,g_2,\dots,g_n,n+1]$, but 
not in the list $[0,g'_1,g'_2,\dots,g'_n,n+1]$.   For example, if 
$g=[1,2,3,4,5], g' =[1,4,3,2,5] \in \Sigma_5$, we have $d_{BP}(g,g')=2$.
It is clear that $\max\{d_{BP}(g,g) \,:\, g,g' \in \Sigma_n\} = n+1$.

Alternatively, one can consider 
the {\em rearrangment distance} between two genomes,
i.e. the minimal number of operations of
a certain type (such as 
transpositions or reversals) that can be applied 
to one of the genomes 
to obtain the other \cite{F09}. 
In terms of Cayley graphs, this distance can be
conveniently expressed for transpositions and reversals
as follows.
Let $$T = T_n := \{t_{i,j} \in  \Sigma_n \,:\, 1 \le i < j \le n \},$$
$$C = C_n := \{ t_{i,i+1} \in T \,:\, 1 \le i \le n-1\},$$ 
(the  {\em Coxeter} generators),
and $$R := \{r_{i,j} \in  \Sigma_n \,:\, 1 \le i < j \le n \}.$$ Note that
all three of these sets generate $\Sigma_n$ \cite{cay} and 
that they are all symmetric, since each generator is its own inverse.
The metric $d_S$, $S = T, C, R$, is
precisely the rearrangement distance.

The diameters of $Cay(\Sigma_n,T)$ and $Cay(\Sigma_n,R)$ 
are both $n-1$, and the diameter of $Cay(\Sigma_n,C)$ 
is  ${n \choose 2}$ \cite{cay}.

Regarding the quantities $\lambda_m(\Sigma_n,S)$, 
we have the following result for $S = T, C, R$:

\begin{theorem}\label{bounds}
For $n \geq 7$ the following hold:
\begin{itemize}
\item[(i)] $\lambda_1(\Sigma_n,T_n) = 1$ and $\lambda_2(\Sigma_n,T_n) = 2$.

\item[(ii)] $\lambda_1(\Sigma_n,C_n)=2n-3$ 
and $2n-2\leq \lambda_2(\Sigma_n,C_n) \leq 4n-6$.

\item[(iii)] $\frac{n+1}{2} \le \lambda_m(\Sigma_n,R_n) \le n-1$, $m=1,2$.
\end{itemize}
\end{theorem}
{\em Proof:}
(i) Note that if $g \in \Sigma_n$ and $t_{i,j} \in T$, then:
\begin{equation}\label{transpose} 
g^{-1} t_{i,j} g = t_{g^{-1}(i),g^{-1}(j)}. 
\end{equation}
Therefore $\lambda_1(\Sigma_n,T) = 1$ by (\ref{l1eq}).
Thus,  by Inequality~(\ref{lemlam}), we have $\lambda_2(\Sigma_n,T) \le 2$. The 
equality $\lambda_2(\Sigma_n,T) = 2$ follows by (\ref{l2eq}) and the
fact that $g^{-1} t_{k,l} t_{i,j} g = 
t_{g^{-1}(i),g^{-1}(j)} t_{g^{-1}(k),g^{-1}(l)}$ holds 
for any $g\in \Sigma_n$ and  $1 \le i < j < k < l \le n$.

(ii) Consider the permutation $g \in \Sigma_n$ given by 
$g = [2,3,\dots,n-1,n,1]$.
Then $g^{-1} t_{1,2}g = [n,2,3,\dots,n-1,1]$. 
Therefore, $l_C(g^{-1}t_{1,2}g) \ge  2n-3$
(since to transform $[n,2,3,\dots,n-1,1]$ to $1_{\Sigma_n}$ requires
moving $1$ and $n$ back to their original positions). 
Therefore,  $\lambda_1(\Sigma_n,C) \ge 2n-3$ by  (\ref{l1eq}).
But, by Equality (\ref{transpose}), $\lambda_1(\Sigma_n,C) \le 2n-3$, 
since any transposition is the product of at most $2n-3$ elements in $C$.
In particular, $\lambda_1(\Sigma_n,C) = 2n-3$.

Similarly, $l_C(g^{-1}t_{1,2}t_{3,4}g) \ge 2n-2$, and 
so $\lambda_2(\Sigma_n,C) \ge 2n-2$ by (\ref{l2eq}).
Hence, by Inequality (\ref{lemlam}), we 
have $\lambda_2(\Sigma_n,C) \le 2(2n-3)$.

(iii) The inequality $\lambda_m(\Sigma_n,R_n) \le n-1$, $m=1,2$ follows
as the diameter of $Cay(\Sigma_n,R)$ is at most $n-1$. 

Now, suppose $n$ is odd. Let $g \in \Sigma_n$ be given by  
$g = [3,2,5,4,7,6,\dots,n-3,n,n-1,1]$.
Then it is
straight-forward to check that $d_{BP}(r_{1,n} g,g) = n+1$
(see Figure~2 for the case $n=7$).
In particular, since the length of any shortest 
path in $Cay(\Sigma_n,R)$ joining any $g,h \in \Sigma_n$ is at least 
$d_{BP}(h,g)/2$ by \cite[p.238]{SM97},
we have $\lambda_1(\Sigma_n,R) \ge \frac{n+1}{2}$. 
Similarly, $d_{BP}(r_{2,3} r_{1,n} g, g) = n+1$ 
for any $g \in \Sigma_n$, and so 
$\lambda_2(\Sigma_n,R) \ge \frac{n+1}{2}$.

\begin{figure}[t]
\label{perms}\begin{center}
 \resizebox{8cm}{!}{
 \includegraphics{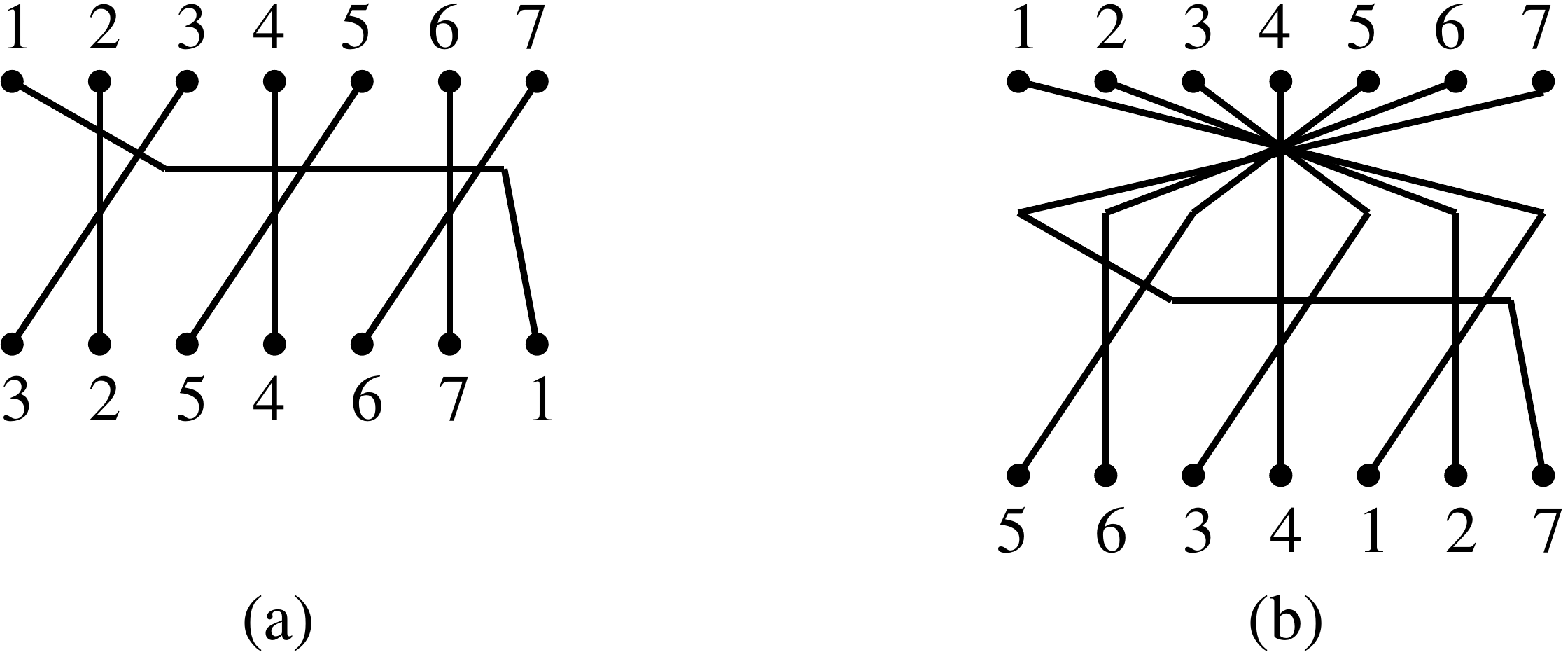}
 }
\end{center}
\caption{(a) A diagrammatic respresentation
of the element $g=[3,2,5,4,6,7,1]$ in $\Sigma_7$, 
defined in the proof of Theorem~\ref{bounds} (iii). (b) The
product $r_{1,7}g = [5,6,3,4,1,2,7]$. Note that 
$d_{BP}(r_{1,7}g,g)=8$.}
\end{figure}

In case $n$ is even, consider $g =  [3,2,5,4,7,6,\dots,n-4,n-1,n-2,1,n]$.
Then $d_{BP}(r_{2,n}g,g) = n+1$ and $d_{BP}(r_{3,4} r_{2,n} g,g) = n+1$. 
Similar reasoning yields the desired result.
\hfill$\Box$

In genomics, the direction in which a gene is oriented in
a genome can also provide useful information to incorporate
in rearrangement models, which 
can be expressed as follows in terms of Cayley graphs \cite{cay}. 
The {\em hyperoctahedral group} $B_n$ 
is defined as the group of all permutations $g^{\sigma}$
acting on the set $\{\pm1,\dots,\pm n\}$ such that 
$g^{\sigma}(-i) = -g^{\sigma}(i)$
for all $i \in [n]$. An element of $B_n$ is a 
{\em signed permutation}.
Signed versions of transpositions and reversals can be 
defined in the obvious way; a sign change transposition $t^{\sigma}_{i,j}$
switches the values in the $i$th and $j$th positions of a signed permutation
as well as both of their signs and so forth. Note
that we also allow $i=j$ for signed transpositions and reversals 
so that $t_{i,i} = r_{i,i}$, $i \in [n]$,  
simply switches the sign of the $i$th value. 
We denote the set of signed 
elements corresponding to those in $S =T,C,R$, 
together with the elements $t_{i,i}, 1\le i \le n$,  
by $S^{\sigma}$. Note that the 
diameter of $Cay(B_n,R^{\sigma})$ is $n+1$ \cite{cay}.

Now, regarding the group $B_n$ as a wreath product \cite[p. 2756]{cay}, 
we have a short exact sequence:
\begin{equation}
\label{fortunate}
1 \rightarrow N \rightarrow B_n   \buildrel{p}\over{\rightarrow} \Sigma_n \rightarrow 1,
\end{equation}
where the homomorphism $p: B_n \rightarrow \Sigma_n$ 
sends $g^{\sigma} \in B_n$ to
the permutation of $[n]$ that maps $i$ 
to $|g^{\sigma}(i)|$ (i.e. it ignores the sign).
Notice that $p$ maps $S^{\sigma}$ onto $S$ when $S=T,C,R$.
In particular, from Lemma~\ref{lower},  the 
following holds for $m=1,2$:

\begin{equation}
\label{helpsus}
\lambda_m(B_n,S^{\sigma}) \ge \lambda_m(\Sigma_n,S).
\end{equation}

Moreover, $N = Ker(p)$ is isomorphic to the 
elementary Abelian 2-group $\ZZ_2^n$ and the 
short exact sequence in (\ref{fortunate}) splits, so $B_n$ is a semidirect product of $\ZZ_2^n$ 
and a subgroup isomorphic to $\Sigma_n$.
Using these observations, we obtain:

\begin{corollary}
For $n \geq 7$, the following hold:
\begin{itemize}
\item[(i)] 
$\lambda_1(B_n, T_n^{\sigma}) \le 3$ and 
$\lambda_2(B_n,T_n^{\sigma}) \le 6$.

\item[(ii)] $2n-3 \le \lambda_1(B_n,C_n^{\sigma}) \le 2n-1$ and 
$2n-2 \le \lambda_2(B_n,C_n^{\sigma})  \leq 4n-2$.

\item[(iii)] $\frac{n+1}{2} \le 
\lambda_m(B_n,R_n^{\sigma}) \le n+1$, $m=1,2$.
\end{itemize}
\end{corollary}
{\em Proof:}
The inequalities $\lambda_1(B_n, T_n^{\sigma}) \le 3$ and
$\lambda_1(B_n,C_n^{\sigma}) \le 2n-1$ follow from
similar arguments to those used in the proof of 
Theorem~\ref{bounds} (i) and (ii), using the 
signed analogue of Equation~(\ref{transpose}). 
Inequality~(\ref{lemlam}) then implies that
inequalities $\lambda_2(B_n,T_n^{\sigma}) \le 6$ and 
$\lambda_2(B_n,C_n^{\sigma})  \leq 4n-2$ both hold.
The inequality $\lambda_m(B_n,R^{\sigma}_n) \le n+1$, $m=1,2$, follows
as the diameter of $Cay(B_n,R^{\sigma}_n)$ is at most $n+1$. 
The inequalities $2n-3 \le \lambda_1(B_n,C_n^{\sigma})$ and
$2n-2 \le \lambda_2(B_n,C_n^{\sigma})$,
and the remaining ones in (iii) follow by Inequality (\ref{helpsus}) 
and Theorem~\ref{bounds}.
\hfill$\Box$

\section{Beyond $d_S$: properties of breakpoint distance}\label{break}

As we have seen for the breakpoint distance on $\Sigma_n$ in the last 
section, it can sometimes be useful to consider 
metrics on a group other than the distance $d_S$ 
arising from some Cayley graph.
Motivated by this, given an arbitrary metric $d$ on a  
finite group $G$, with symmetric generator set $S$, we define:
$$
\lambda_1(G,S,d): = \mathop{\displaystyle \max }_{\scriptscriptstyle g \in G, s \in S}\limits^{} \{d(sg, g)\} \mbox{ and } 
\lambda_2(G,S,d): =  \mathop{\displaystyle \max }_{\scriptscriptstyle g \in G, s,s' \in S}\limits^{} \{d(sg, s'g)\}.
$$

In particular, $\lambda_m(G,S) = \lambda_m(G,S,d_S)$
and $\lambda_m(G,S,d) \leq \max_{g,g' \in G}\{d(g,g')\}$, $m=1,2$.
Moreover, the following analogue of Inequality~(\ref{lemlam})
for an arbitrary metric $d$ on $G$  is easily seen to hold: 
\begin{equation}\label{analogue}
\lambda_2(G,S,d) \leq 2 \cdot \lambda_1(G,S,d).
\end{equation}

Note that, although the quantities $\lambda_m(G,S)$ and $\lambda_m(G,S,d)$ 
need not be directly related to one another, in certain circumstances,
they are. For example, if $d$ has the property that $d(g, gs) \leq c$ 
for some constant $c$ it is an easy exercise to show 
that $\lambda_m(G,S,d) \leq c \cdot \lambda_m(G,S),$ for $m=1,2$.

We now return to considering the breakpoint distance $d_{BP}$.
In genomics, this distance 
is commonly used as a proxy for rearrangement distances.
Thus it is of interest to note:

\begin{lemma}\label{bp}
For $n \geq 7$,  the following hold:
\begin{itemize}
\item[(i)] $\lambda_1(\Sigma_n,T_n,d_{BP}) \le 4$ and $\lambda_2(\Sigma_n,T_n,d_{BP}) \le 8$.

\item[(ii)] $\lambda_1(\Sigma_n,C_n,d_{BP})\le 4$ and $\lambda_2(\Sigma_n,C_n,d_{BP})  \le 8$.

\item[(iii)] $\frac{n+1}{2} \le 
\lambda_m(\Sigma_n,R_n,d_{BP}) \le n+1$, $m=1,2$.
\end{itemize}
\end{lemma}
{\em Proof:}
Suppose $t=t_{i,j} \in T_n$, $1 \le i < j \le n$.  
Using Equation~(\ref{transpose}), it is straightforward 
to see that  $d_{BP}(tg,g) \leq 4$ holds for any $g \in \Sigma_n$.
Therefore $\lambda_1(\Sigma_n,T_n,d_{BP}), 
\lambda_1(\Sigma_n,C_n,d_{BP}) \le 4$.
The inequalities in (i) and (ii) involving $\lambda_2$ 
now follow from Inequality (\ref{analogue}).

The Inequalities in (iii) follow from the 
argument used in the proof of
Theorem~\ref{bounds} (iii) and the diameter of $d_{BP}$ on $\Sigma_n$.
\hfill$\Box$
 
In particular, for $C$, the set of Coxeter generators of $\Sigma_n$
in the last section, and $m=1,2$, we have  $\lambda_m(\Sigma_n,C) \ge 2n-3$,
but $\lambda_m(\Sigma_n,C,d_{BP}) \le 4$. Intriguingly, 
this observation can be extended as follows. 
For $k \ge 1$, let $R^{(k)}$, denote the set
of reversals of the form $\{r_{i,j} \,:\, 1 \le i < j \le n, |i-j| \le k\}$.
Such `fixed-length' reversals have been 
considered in the context of genome 
rearrangements in e.g. \cite{CS96}. 
Note that
$R^{(1)}=C$ and $R^{(k)} \subseteq R^{(k+1)}$, 
so that $R^{(k)}$ generates $\Sigma_n$.

\begin{proposition}
For $n \ge 7$, $n \ge k \ge 1$ and $m =1,2$,  
$$\lambda_m(\Sigma_n,R^{(k)}) \ge 2 \lceil \frac{n}{k} \rceil-2,$$  
and
$$\lambda_m(\Sigma_n,R^{(k)},d_{BP}) \le 4(k+1).$$
\end{proposition}
{\em Proof:}
As in the proof of Theorem~\ref{bounds} (ii), let $g \in \Sigma_n$ 
be given by $g = [2,3,\dots,n-1,n,1]$, 
so that $g^{-1}r_{1,2}g = [n,2,3,\dots,n-1,1]$. 
Then, $l_{R^{(k)}}(g^{-1}r_{1,2}g) \ge 2 \lceil \frac{n}{k}\rceil-3$, 
since to transform $[n,2,3,\dots,1]$ to $1_{\Sigma_n}$ requires
moving $1$ and $n$ back to their original positions. 
Similarly, $l_C(g^{-1}r_{1,2}r_{3,4}g) \ge 2 \lceil \frac{n}{k}\rceil-2$.
This gives the first inequality in the proposition.
Moreover, if $r_{i,j}, r_{p,q} \in R^{(k)}$, then 
it is straight-forward to see 
that $d_{BP}(r_{i,j}g,g) \le 2(k+1)$ and  
$d_{BP}(r_{p,q}r_{i,j}g,g) \le 4(k+1)$ holds, which gives 
the second inequality in the proposition.
\hfill$\Box$

This proposition 
implies that in genomics applications, adding or 
substituting a single reversal in a sequence of reversals in $R^{(k)}$ 
could potentially have a large effect on $d_{R^{(k)}}$, but a 
relatively small effect on $d_{BP}$ (especially 
for large values of $n$, e.g. there are $n \ge 20,000$ genes in the human genome).
It could be of interest to see whether other combinations of 
generating sets and metrics for $\Sigma_n$
commonly used in genomics (such as  transpositions \cite{L06} 
and the $k$-mer distance \cite{TR10}) exhibit a similar 
type of behaviour. 

\section{Statistical implications}
\label{stats}

So far we have considered metric sensitivity from a purely combinatorial and deterministic perspective.  But it is also of interest to investigate the sensitivity of the metrics discussed above when the
elements of $S$ are randomly assigned.  Again, the motivation for this question comes from genomics, where stochastic models often play a central role (see, for example, 
\cite{mos}, \cite{Wang}).  In this section, we establish a result (Proposition \ref{random}) in which the quantity $\lambda_2$ plays a crucial role in allowing underlying parameters in such stochastic models to be estimated accurately given sufficiently long genome sequences. Our motivation here is to provide some basis for eventually extending the well-developed (and tight) results on the sequence length requirements for tree reconstruction under site-substitution models (see e.g. \cite{das, erd, gro, mos}) to more general models of genome evolution.

Consider any model of genome evolution, where an associated transformation group $G$ acts freely on a set $X$ of genomes of length $n$, and for which events in  some symmetric generating set $S$ occur independently according to a Poisson process.  Regard the elements of $X$ as leaves of an evolutionary (phylogenetic) tree
with weighted edges \cite{sem}, and let $\mu(x,y)$ be the sum of the weights of the edges of the tree connecting leaves $x, y$.  Then we make the following assumption:

\begin{itemize}
\item
The expected number of times that $s \in S$ occurs along the path in the tree connecting $x$ and $y$ can be written as $n \cdot \mu_s(x,y)$  (i.e. we assume that the rate of events scales linearly with the length of the genome).
\end{itemize}

Let $\mu(x,y)= \sum_{s \in S} \mu_s(x,y)$. Then the total number of events in $S$ that occur on the path separating $x$ and $y$ has a Poisson distribution with mean $n \cdot \mu(x,y)$.  

Now suppose $d$ is some metric on genomes  that satisfies the following three properties:

\begin{itemize}
\item[(i)] $d(x,g\circ x)$ depends just on $g$, for each  $x \in X$ and $g \in G$.
\item[(ii)] $\lambda_2(G,S, d)$ is independent of $n$.
\item[(iii)] $\overline{d} =  nf(\mu(x,y)),$
where $\overline{d}$ is the expected value in the model of $d(x,y)$  and 
$f$ is a function with strictly positive but bounded first derivative on $(0, \infty)$.
\end{itemize}

An example to illustrate this process is site substitutions, under the Kimura 3ST model, described at the start of Section \ref{genome}, taking $d=d_S$, where we observed that Properties (i) and (ii)  hold (note that in this case, $d(x,y)$ is the `Hamming distance' between the sequences which counts the number of sites at which $x$ and $y$ differ).
In that case, Property (iii) also holds, since
$$\overline{d} = n \frac{3}{4}(1- \exp(-4\mu(x,y)/3)).$$
Note that, both breakpoint distance and $d_S$ satisfy (i), and we have described above some cases where (ii) is satisfied.  Whether (iii) holds (or the assumption that the expected number of events scales linearly with $n$) depends on the 
details of the underlying stochastic process of genome rearrangement.  For example, for the approximation to the Nadeau-Taylor model of genome rearrangement studied in Section 2 of  \cite{wan}, Property (iii) holds under the assumption that the number of events separating $x$ and $y$ has a Poisson distribution whose mean scales linearly with $n$ (the proof  relies on Corollary 1(a) of \cite{wan}).

The following  result shows how $d/n$ can be used to estimate $f(\mu(x,y))$ accurately, and thereby $\mu(x,y)$ (by the assumptions regarding $f$).  The ability to estimate $\mu(x,y)$ accurately provides a direct route to accurate tree reconstruction by standard phylogenetic methods 
(such as `neighbor-joining' \cite{sn}) since $\mu(x,y)$ is `additive' on the underlying tree but not on alternative binary trees (for details, see \cite{sem}). 

\begin{proposition}
\label{random}
Consider any stochastic model of genome evolution  for which events in $S$ occur according to a Poisson process with a rate that scales linearly with $n$, and any metric
$d$ that satisfies conditions (i) --(iii) above. Then the probability that $d(x,y)/n$ differs from $f(\mu(x,y))$ by more than $z$ converges to zero exponentially quickly with increasing $n$. 
More precisely,  for constants $b>0$ and $c>0$ that depend just on $\mu(x,y)$ and on the pair  $(\lambda_2(G,S,d), \mu(x,y)$), respectively, we have:
$$\PP(|d/n - f(\mu(x,y))| \geq  z)  \leq \exp(-bn) + 2 \exp(-cz^2n),$$ 
for $d=d(x,y)$.
\end{proposition}

\noindent {\em Proof of Proposition \ref{random}:}
We first recall the Azuma-Hoeffding inequality (see e.g. \cite{Alon})  in which  $X_1, X_2, \ldots, X_k$ are independent random variables taking values in some set $S$, and $h$ is 
{\em any} real-valued function defined on $S$ that satisfies the following property for some constant $\xi$:
$$|h(x_1, x_2, \ldots, x_k) - h(x'_x, x'_2, \ldots, x'_k)| \leq \xi,$$ whenever $(x_i)$ and $(x'_i)$ differ at just one coordinate. In this case,  the random variable
$Y:= h(X_1, X_2, \ldots, X_k)$ has the tight concentration bound for all $k>1$:
\begin{equation}
\label{azuma}
\PP(|Y - \EE[Y]| \geq z) \leq 2\exp(-\frac{z^2}{2\xi^2k}).
\end{equation}
We apply this general result as follows.
Let $K$ be the random total number of events in $S$ that occur in the path separating $x$ and $y$. By assumption, $K$
 has a Poisson distribution with mean $n \cdot  \mu(x,y)$. Conditional on the event $K=k$,  let $X_1,\ldots, X_k$ be the actual elements of $S$ that 
 occur.  It is assumed that these events are independent. Moreover, by (i), $d(x,y)$ is a function of $X_1,\ldots, X_k$, and by (ii) this function satisfies the requirements of the Azuma-Hoeffding inequality for $\xi =\lambda_2(G,S,d)$. Thus  (\ref{azuma}) furnishes the following inequality:
 \begin{equation}
 \label{azh}
 \PP(|d/n - \overline{d}/n| \geq  z \, | \, K=k) \leq 2\exp(-\frac{z^2n^2}{2\lambda^2k}).
 \end{equation}
Invoking Property (iii) and the law of total probability, we obtain:
$$\PP(|d/n - f(\mu(x,y))| \geq  z) = \sum_{k\geq 0} \PP(|d/n - \overline{d}/n| \geq z \, | \, K=k)\PP(K=k),$$
from which (\ref{azh}) ensures the inequality:
 \begin{equation}
 \label{pp1}
 \PP(|d/n - f(\mu(x,y))| \geq  z)  \leq 2\EE[\exp(-\frac{z^2n^2}{2\lambda^2K})],
 \end{equation}
 where $\EE$ denotes expectation with respect to $K$. 
Let us write $\EE[\exp(-\frac{z^2n^2}{2\lambda^2K})]$ as a weighted sum of two conditional expectations:
 \begin{equation}
 \label{pp2}
\EE[\exp(-\frac{z^2n^2}{2\lambda^2K})|K > 2n \cdot  \mu(x,y)]\cdot p +\\
 \EE[\exp(-\frac{z^2n^2}{2\lambda^2K})|K \leq 2n \cdot  \mu(x,y)] \cdot(1-p), 
 \end{equation}
 where $p =  \PP(K > 2n \cdot  \mu(x,y))$.  The first term in (\ref{pp2}) is bounded above by $\PP(K > 2n \cdot  \mu(x,y))$ since $\exp(-\frac{z^2n^2}{2\lambda^2K}) \leq 1$; moreover,  since $K$ has a Poisson distribution with mean $n \cdot  \mu(x,y)$ (and so  is asymptotically  normally distributed with mean and variance equal to $\mu n$), the quantity $\PP(K > 2n \cdot  \mu(x,y))$ is bounded above by a term of the form $\exp(-bn)$ where $b$ depends just on $\mu(x,y)$. 
 
The second term in (\ref{pp2})
 is bounded above by $\exp(-\frac{z^2n}{4\lambda^2\mu(x,y)})$,
 where $\lambda = \lambda_2(G,S,d)$, since the function $x \mapsto \exp(-A/x)$ increases monotonically on $[0, \infty)$. 
 
 Combining these two bounds in (\ref{pp2}), the result now follows from (\ref{pp1}).

 \hfill $\Box$ \\

\noindent {\bf Remark.}  Referring again to the particular case of site substitutions under the Kimura 3ST model, Proposition~\ref{random} can be strengthened to: 
$$\PP(|d/n - f(\mu(x,y))| \geq  z)  \leq \ 2 \exp(-c'z^2n),$$ where $c'>0$ can be chosen to be independent of $\mu(x,y)$.
This  stronger result is the basis of numerous results in the phylogenetic literature that
show that large trees can be reconstructed from remarkably short sequences under simple site-substitution models \cite{erd}.
Although the bound in Proposition \ref{random} is less incisive, 
it would be of interest to explore similar phylogenetic applications for other models of genome evolution in which $\lambda_2$ is independent of $n$,
such as those involving breakpoint distance under reversals of fixed length. \\

\bigskip

\noindent{ \bf Acknowledgments}

\bigskip

We thank Marston Conder, Eamonn O'Brien and Li San Wang for some helpful comments.
VM thanks the Royal Society for supporting his visit to University 
of Canterbury, where most of this work was undertaken. MS thanks the Royal Society of New Zealand under its
James Cook Fellowship scheme.

\newpage


\bibliographystyle{elsarticle-num}

\end{document}